# Mid-infrared wide-field nanoscopy

Miu Tamamitsu[1], Keiichiro Toda[1], Venkata Ramaiah Badarla[1], Hiroyuki Shimada[1], Kuniaki Konishi[1], and Takuro Ideguchi[1,*]

[1]Institute for Photon Science and Technology, The University of Tokyo, Tokyo 113-0033, Japan

*ideguchi@ipst.s.u-tokyo.ac.jp

**Abstract**

Mid-infrared (MIR) spectroscopy is widely recognized as a powerful, non-distractive method for chemical analysis. However, its utility is constrained by a micrometer-scale spatial resolution imposed by the long-wavelength MIR diffraction limit. This limitation has been recently overcome by MIR photothermal (MIP) imaging, which detects photothermal effects induced in the vicinity of MIR absorbers using a visible-light microscope. Despite its promise, the full potential of its spatial resolving power has not been realized. Here, we present an optimal implementation of wide-field MIP imaging to achieve high spatial resolution. This is accomplished by employing single-objective synthetic-aperture quantitative phase imaging (SOSA-QPI) with synchronized sub-nanosecond MIR and visible light sources, effectively suppressing the resolution-degradation effect caused by photothermal heat diffusion. We demonstrate far-field MIR spectroscopic imaging with a spatial resolution limited by the visible diffraction, down to 125 nm, in the MIR region of 3.12–3.85 μm (2,600–3,200 cm$^{-1}$). This technique, through the use of a shorter visible wavelength and/or a higher objective numerical aperture, holds the potential to achieve a spatial resolution of less than 100 nm, thus paving the way for MIR wide-field nanoscopy.

**Introduction**

Vibrational spectroscopy, including mid-infrared (MIR) and Raman spectroscopy, is one of the gold standards for label-free and non-destructive chemical analysis. While MIR spectroscopy has the advantage of higher sensitivity than Raman spectroscopy, one of its critical drawbacks is the low spatial resolution of a few micrometers imposed by the diffraction limit of the long MIR wavelength. To overcome this fundamental barrier, atomic force microscopy has been used to sense the MIR-absorption-induced temperature rise or thermal expansion of the sample with its nanoscale probe[1]. This method can realize a spatial resolution as low as a few tens of nanometers, finding applications in characterizing various types of nanoscale materials such as polymer films and plasmonic structures. However, it is not applicable to certain types of samples, such as living biological specimens, due to the near-field operation. To address this limitation, various far-field sub-diffraction-limited MIR imaging methods have been recently developed, including vibrationally resonant nonlinear imaging[2,3], ultraviolet (UV)-localized MIR photoacoustic imaging[4], and MIR photothermal (MIP) imaging[5]. These methods have enabled MIR imaging of submicron structures of living cells and pathology slides, as well as the detection of single viruses and bacteria.

Despite their advances, there has been difficulty in achieving a spatial resolution of less than 100 nm, a challenging milestone for far-field optical imaging that distinguishes nanoscopy from microscopy[6]. One of the limiting factors is the lack of a high numerical aperture (NA > 1) objective lens in the UV or MIR region. For example, in vibrationally resonant nonlinear imaging[2,3], MIR and either near-infrared (NIR) or visible (VIS) beams are used to excite the sample's nonlinear scattering. However, to achieve a high spatial resolution, the technique requires collinear focusing of MIR and NIR/VIS beams onto the sample. This requirement calls for a MIR-compatible objective lens for which a high NA is not available (i.e., NAs are typically below ~0.65), consequently limiting the spatial resolution to ~450 nm in full width at half maximum (FWHM)[3]. In UV-localized MIR photoacoustic imaging[4], the change in the Grüneisen parameter upon MIR-absorption-induced temperature rise is imaged by UV photoacoustic imaging. In this technique, the use of UV light is essential to ensure its applicability to various types of samples, as the UV light is absorbed by various molecular structures to generate a photoacoustic signal. However, the lack of a high-NA objective lens in the UV region (e.g., NA ~0.45) prohibits this technique from fully exploiting the advantage of the short UV wavelength, thus limiting its spatial resolution to ~260 nm in FWHM.

In contrast, MIP imaging[5] offers the potential to achieve higher spatial resolution in far-field MIR imaging because its detection is based on the VIS light, for which high-NA objective lenses are readily available. However, conventional implementations are suboptimal and often hindered by the heat diffusion of the photothermal effect, leading to a degradation of spatial resolution. For instance, earlier demonstrations of MIP imaging employed continuous-wave (CW) VIS light to detect the photothermal signal induced by pulsed or CW MIR light, both in point-scanning and wide-field configurations[7-12]. In these cases, the spatial resolution is limited by the characteristic length of heat diffusion, typically several hundred nanometers[13]. Recently, a spatial resolution of 120 nm was demonstrated by higher-order harmonic lock-in detection, which mitigates the heat diffusion problem in the point-scanning configuration[14]. However, taking higher-order harmonics significantly decreases the signal intensity by an order of magnitude, resulting in a relatively long pixel dwell time of ~200 ms. Furthermore, the point-scanning method suffers from limitations in imaging speed due to the slow mechanical scan of the sample stage. These issues can be addressed by employing a pump-probe wide-field configuration that uses ns pulses and an image sensor. In this method, a MIR pulse excites the molecular vibrations in a wide-field manner, and the induced photothermal heat is subsequently probed by a VIS pulse and captured with an image sensor, ideally before noticeable heat diffusion occurs[15-21]. Although video-rate imaging has been demonstrated with this method[21], previous demonstrations have not realized its full spatial resolving power due to the use of air objective lenses with moderate NAs and pump-probe pulses with durations of more than 10 ns, which degraded the spatial resolution by more than 100 nm[22] in an aqueous environment.

In this work, we present a wide-field MIP imaging technique optimized for high spatial resolution, paving the way for far-field MIR nanoscopy. To prevent degradation of spatial resolution due to the photothermal heat diffusion, we developed passively-synchronized sub-nanosecond MIR and VIS light sources. These sources were generated through nonlinear wavelength conversions from a single NIR seed laser. To enable the use of a high-NA immersion objective lens for VIS imaging while simultaneously illuminating the sample with the MIR light, we developed a single-objective synthetic-aperture quantitative phase imaging (SOSA-QPI) technique featuring a VIS-reflective and

MIR-transparent sample holder. This system facilitates VIS-diffraction-limited MIP imaging and accomplishes far-field MIR spectroscopic imaging with an exceptional spatial resolution of 125 nm within the MIR region of 3.12 - 3.85 μm (2,600 - 3,200 cm$^{-1}$), surpassing the MIR diffraction limit by a factor of 30. Moreover, unlike the higher-order lock-in detection scheme[14], in which most of the CW probe photons sense the diffused photothermal signals that do not contribute to the enhancement of spatial resolution, our nanosecond pump-probe scheme ensures optimal use of the probe photons that only sense the non-diffused photothermal signals. This results in an order of magnitude higher imaging speed of less than 1 minute/frame. A straightforward extension for utilizing a shorter VIS wavelength (e.g., ~400 nm) and/or higher objective NA (e.g., ~1.4) can further enhance this technique, leading to a spatial resolution below 100 nm.

**Results**

**Principle of MIP-QPI with aperture synthesis**

We first introduce the concept of MIP imaging with aperture synthesis. In this work, we employ QPI to detect the MIP effect in a wide-field manner (Fig. 1). QPI measures the VIS complex-field image consisting of amplitude and optical-phase-delay (OPD) distributions. Therefore, the MIP-QPI[16,18,19,21] technique quantifies local refractive-index changes occurring in the vicinity of MIR absorbers due to the MIP effect while also revealing the morphology of transparent objects. The MIP contrast is determined by measuring the OPD distributions induced by the sample when the MIR light is turned on and off, followed by taking the difference between these two images. To improve the spatial resolution of QPI, we perform aperture synthesis. In this method, the sample is illuminated by a plane wave with an angle relative to the objective lens' detection plane, which shifts the center position of the objective lens' circular bandpass filter with respect to the sample's spatial frequency spectrum (Fig. 1c). This lets us probe the higher-frequency contents that cannot be accessed with the normal illumination. By scanning the azimuthal angle of the illumination, the position of the shifted bandpass filter is also rotated around the zero-frequency point of the sample's spectrum, facilitating the measurement of different high-frequency regions. Using these multiple images captured under different illumination vectors, we can numerically synthesize a virtual imaging aperture with a larger spatial-frequency bandwidth than the physical bandwidth of the objective lens. This results in a higher spatial resolution in the final reconstructed image.

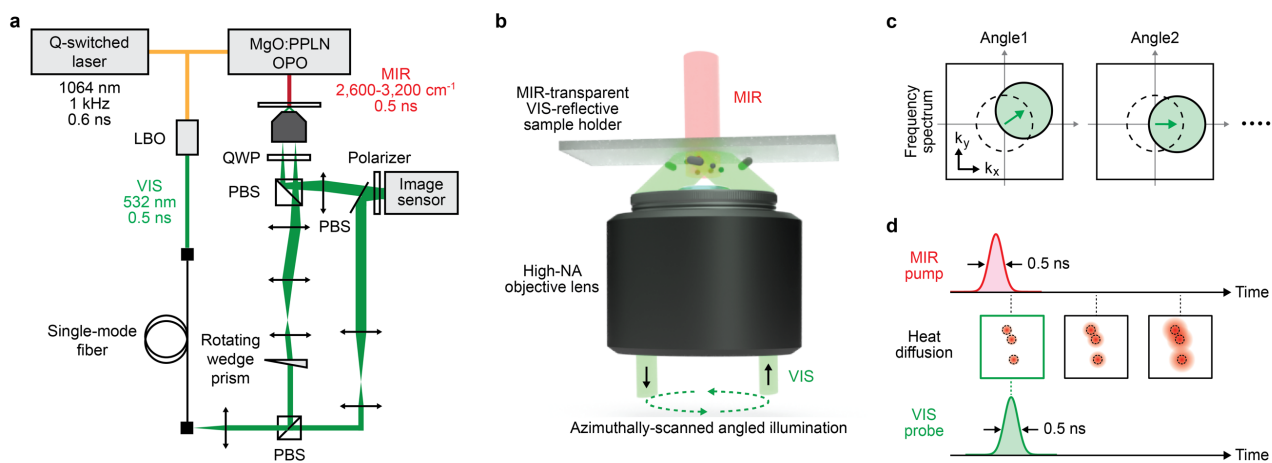

**Figure 1. Single-objective synthetic-aperture (SOSA) MIP-QPI. a** Schematic representation of the system. **b** Detailed depiction of the area around the sample on a VIS-reflective/MIR-transparent sample holder. **c** Regions of the sample's spatial-frequency spectrum probed with the objective lens with angled illuminations. **d** Temporal representation of the pump and probe pulses. OPO: optical parametric oscillator, LBO: lithium triborate, PBS: polarizing beamsplitter, QWP: quarter-wave plate.

## Passively-synchronized sub-nanosecond VIS and MIR light sources

To achieve the optimum spatial resolution and efficient use of the probe photons in MIP imaging, it is crucial to detect the photothermal effect with the VIS light before the induced heat from a MIR absorber diffuses beyond the VIS diffraction limit. Based on our theoretical considerations, we estimate that MIR and VIS pulse durations of ~1 ns meet this requirement when the VIS spatial resolution is approximately 100 nm (see Methods for details). To realize accurate synchronization between the MIR and VIS pulses, we developed passively-synchronized MIR and VIS light sources through nonlinear wavelength conversions from a single Q-switched Nd: YAG laser emitting 1,064-nm, 0.65-ns pulses at 1 kHz (Fig. 1a). The VIS pulses are obtained through second harmonic generation using a 30-mm-long lithium triborate (LBO) crystal. The resultant 532-nm pulses are then coupled into a single-mode optical fiber. The MIR pulses are produced with our home-made optical parametric oscillator (OPO) based on a 50-mm-long fan-out magnesium-oxide-doped periodically-poled lithium niobate (MgO:PPLN) crystal, allowing for tuning of the idler MIR light in the region of 2,600–3,200 cm$^{-1}$ (3.12–3.85 μm) through mechanical scanning of the crystal's position (see Supplementary Notes 1 and 2 for details). Finally, the output MIR pulse train is intensity modulated at 35.7 Hz (1/28th harmonic of the 1-kHz pulse repetition frequency) by a mechanical chopper. A detailed schematic of the laser system can be found in Supplementary Note 1. We confirmed that the VIS and MIR pulses have sub-nanosecond pulse durations of ~0.5 ns in FWHM (see Supplementary Note 2 for details). The optical path length of the MIR beam is adjusted such that the VIS pulse arrives at the falling edge of the MIR pulse at the sample plane (Fig. 1d). The MIR pulses exhibit pulse energy between 0.6 and 1.7 μJ and spectral bandwidth between 15 and 50 cm$^{-1}$ in FWHM, depending on the wavenumber, and can resolve CH stretching vibrational signatures between 2,700 and 3,100 cm$^{-1}$ (see Supplementary Note 2 for details).

**Single-objective synthetic-aperture (SOSA) MIP-QPI**

To achieve high spatial resolution in MIP images using QPI, we developed a synthetic-aperture quantitative phase microscope (Fig. 1) that allows the use of a high-NA immersion objective lens while ensuring that the MIR pulses reach the sample. To accomplish this, we employ a single-objective QPI geometry with a VIS-reflective and MIR-transparent sample holder, such as a silicon substrate. Figures 1a and 1b provide a schematic illustration of our SOSA-MIP-QPI system. The 532-nm pulses emitted from the single-mode optical fiber are collimated and split into two paths of the Mach-Zehnder interferometer. In the sample arm, a wedge prism is used to deflect the beam, delivering an angled beam to the sample with an illuminating NA of 1.0 via a water-immersion objective lens with an NA of 1.2 (UPlanSApo60x, Olympus). We incrementally rotate the wedge prism by a step of 45 degrees, employing 8 different azimuthal angles of illumination for aperture synthesis. The designed half-pitch spatial resolution with aperture synthesis is 121 nm.

To facilitate the experimental system, we utilize a silicon substrate as the sample holder due to its reflective properties in the VIS region and transparency in the MIR region. This allows us to collect the beam reflected from the silicon substrate, which contains forward-scattered light from the sample, through the same objective lens. Thus there is no need to sandwich the sample between two closely positioned immersion-based objective lenses, which would hinder the delivery of MIR light to the sample. In our system, the MIR light can be delivered through the silicon substrate from the opposite side of the objective lens. The light collected by the objective lens is then separated from the original beam path using a polarizing beam splitter (PBS) in conjugation with a quarter-wave plate (QWP). An intermediate image is formed by a subsequent lens with a magnification rate of 66.7. An additional 4f system installed in the following path further magnifies the image by a factor of 5, resulting in a total magnification rate of 333.3. This high magnification rate helps in detecting more photons per diffraction limit with increased optical illumination, thereby reducing the OPD noise in QPI governed by the optical shot noise.

The sample and reference arms are recombined by a PBS, with a polarizer extracting the same polarization from the two arms of the interferometer. In the reference arm, the beam diameter, the optical path length, and the angle of the reference beam relative to the sample beam are optimized to maximize interferometric visibility. Finally, the interferogram is captured by an image sensor (acA2440-75um, Basler) at a rate of 71.4 Hz (1/14th harmonic of the 1-kHz pulse repetition frequency), allowing for alternate probing of MIR-on and -off states for each illuminating angle. The difference between on-off states reveals the MIP contrast. Each MIP image consists of 135 pixels x 135 pixels with a resolution of 121 nm/pixel and is measured in less than 1 minute to achieve a reasonable signal-to-noise ratio. This measurement time is more than an order of magnitude shorter than that of the conventional state-of-the-art far-field super-resolution MIR imaging technique[14], which requires ~200 ms pixel dwell time, equivalent to an hour for a single image. Details on the SOSA-MIP-QPI reconstruction procedure can be found in Supplementary Note 3.

## VIS-diffraction-limited MIP imaging

We demonstrate that our MIP imaging technique minimizes the effect of heat diffusion, with the spatial resolution determined by the VIS diffraction limit. To evaluate this, we fabricated a spatial-resolution test chart consisting of various sizes of nanometer-scale bars made from an organic resist deposited on a 520 μm-thick silicon substrate. Scanning electron microscopy confirmed the presence of bars with width and spacing ranging from 120 to 170 nm (see Supplementary Note 4 for details). For observation using the water-immersion objective lens, we added a few drops of water and a cover glass on top of the silicon substrate and the test chart.

Figure 2a displays the synthetic-aperture OPD image of the test chart without MIR light. Figure 2b presents the MIP image of the same field of view obtained with the MIR light turned on, with a peak wavenumber of 2,920 cm$^{-1}$ corresponding to the $CH_2$ anti-symmetric stretching mode resonance. Deconvolution between these two images (Figs. 2a and 2b) generates the point-spread function (PSF) of the MIP image relative to the OPD image, as shown in Fig. 2c. This PSF represents the extent of spatial resolution degradation in MIP imaging compared to synthetic-aperture QPI. In the absence of spatial resolution degradation, the 2D PSF becomes equivalent to the inverse Fourier transformation of the synthetic aperture filled with unity, as shown in Fig. 2d. Horizontal and vertical cross-sections of these experimental and ideal PSFs are presented in Fig. 2e, indicating that the MIP image provides nearly the same spatial resolution as synthetic-aperture QPI. The FWHM of the horizontal and vertical cross-sections are degraded by approximately 1 nm or less.

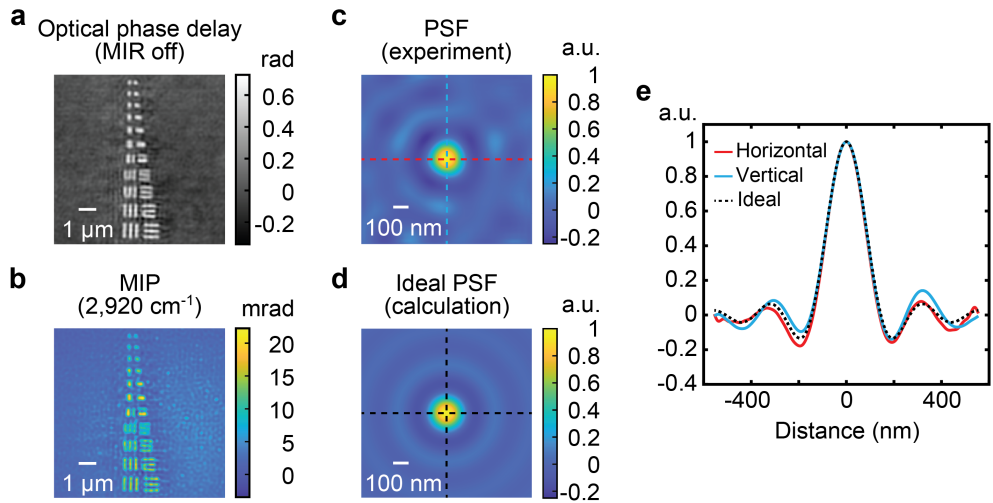

**Figure 2. VIS-diffraction-limited MIP imaging. a** Synthetic-aperture OPD image of the custom-made spatial-resolution test chart with the MIR light turned off. **b** MIP image of the same field of view, captured with the MIR light turned on, which peaks at 2,920 cm$^{-1}$ resonant with the $CH_2$ anti-symmetric stretching mode. **c** Experimental PSF obtained by deconvolving the MIP image **b** by the OPD image **a**. **d** Ideal PSF calculated through inverse Fourier transformation of the synthetic aperture filled with unity. **e** Comparison of the cross-sectional profiles between the experimental and ideal PSFs.

# MIR spectroscopic imaging with 125-nm spatial resolution

We demonstrate MIR spectroscopic imaging at a spatial resolution of 125 nm. To achieve this, we perform hyperspectral MIP-QPI of our custom-made spatial-resolution test chart in the peak wavenumber range between 2,650 and 3,150 cm$^{-1}$, with an increment of 10 cm$^{-1}$. Figure 3a shows a side-by-side comparison of the smallest resolvable set of bars obtained with the MIR light turned off (OPD contrast) and on (MIP contrast at 2,920 or 3,150 cm$^{-1}$). This comparison illustrates that the MIP images possess spatial resolving power nearly equivalent to that of the standard synthetic-aperture QPI. The full-pitch width of the chart is found to be 250 nm, indicating that the system provides a half-pitch spatial resolution of 125 nm. Considering the probe wavelength of 532 nm, this resolution corresponds to a synthetic NA of 2.13, which agrees well with the designed value of 2.2 (i.e., illuminating and collecting NAs of 1.0 and 1.2, respectively).

Figure 3b presents the broadband MIP spectra obtained from one of the bars and the background water, clearly visualizing the difference in vibrational resonance between the two materials. This allows us to estimate that the positive contrast of the test chart in the 2,920-cm$^{-1}$ MIP image (Fig. 3a) corresponds to the vibrational resonance of the $CH_2$ stretching modes present in the organic resist, while its negative contrast in the 3,150-cm$^{-1}$ MIP image (Fig. 3a) reflects the stronger vibrational resonance of the background water. By employing the multivariate curve resolution-alternating least squares (MCR-ALS) algorithm[23], we can decompose the spectral image dataset into two meaningful spectral components, as shown in Figs. 3c and 3d. The retrieved MCR spectra displayed in Fig. 3e indicate that the component labeled as "MCR1" represents the vibrational signature of C-H stretching modes, featuring two characteristic peaks at 2,850 and 2,920 cm$^{-1}$. On the other hand, the other component, labeled as "MCR2," represents the vibrational signature of water, exhibiting increasing vibrational resonance in the higher wavenumber region. Furthermore, the enlarged insets in Figs. 3c and 3d show that the 125-nm bars remain resolved in the two MCR images.

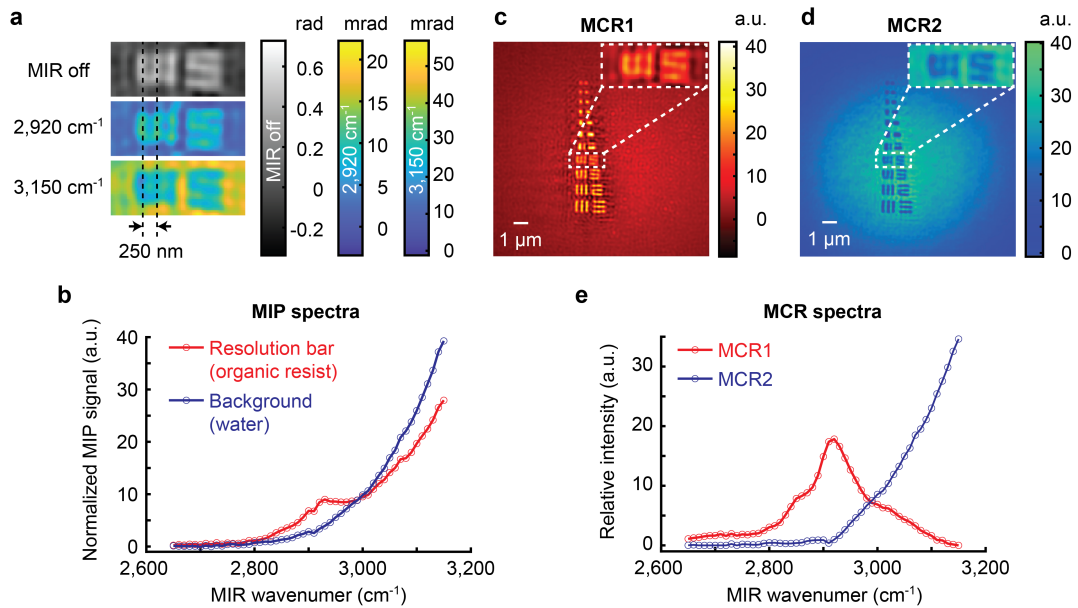

**Figure 3. MIR spectroscopic imaging with 125-nm spatial resolution. a** Smallest resolvable set of bars obtained

with the MIR light turned off (OPD contrast) and on (MIP contrast at 2,920 or 3,150 cm$^{-1}$). Bars with a half-pitch width of 125 nm are resolved in both OPD and MIP images. **b** Broadband MIP spectra obtained on the bars (red) and the background water (blue), displaying distinct vibrational signatures. To derive the normalized MIP signals, each MIP image is divided by the relative MIR pulse energy measured at the corresponding wavenumber. **c-e** MCR-ALS analysis of the broadband MIP spectral image dataset. **c, d** Retrieved images of the two MCR components, exhibiting signatures characteristic of **c** the bars and **d** the background water. The 125-nm bars remain resolved in these images. **e** Retrieved spectra of the two MCR components. The spectrum of MCR1 clearly exhibits the vibrational signature of C-H stretching modes with peaks at 2,850 and 2,920 cm$^{-1}$, while the spectrum of MCR2 displays increasing vibrational resonance in the higher wavenumber region, a characteristic feature of water.

**MIR spectroscopic imaging of bacterial subcellular structures**

We demonstrate the biological applicability of our system by showcasing far-field MIR spectroscopic imaging of bacterial subcellular structures. We perform MIP spectroscopic imaging of a bacterium extracted from commercial fermented soybeans in the peak wavenumber range between 2,800 and 3,500 cm$^{-1}$, with an increment of 20 cm$^{-1}$. Figure 4a presents an example of an OPD image of the bacterium. Through principal component analysis of our raw spectral images, we identify several subcellular structures exhibiting unique MIR spectroscopic contrasts (see Supplementary Note 5 for details). Each of these structures is represented as a binary mask in Figs. 4b-d.

To perform MIR spectroscopic analysis of these subcellular structures, we display averaged and normalized MIP spectra obtained within these binary masks in Fig. 4e. For instance, in the circular structure represented by Fig. 4b, two sharp spectral peaks are observed at 2,840 and 2,920 cm$^{-1}$. These are likely attributed to the symmetric and anti-symmetric stretching modes of CH$_2$ functional groups, suggesting the presence of a lipid droplet. In the membrane-like structure presented in Fig. 4c, a spectral peak, characteristic of the anti-symmetric stretching mode of CH$_3$ functional groups, appears at 2,960 cm$^{-1}$. Additionally, the spectrum exhibits broad peaks at 3,080 and 3,280 cm$^{-1}$, hypothesized to be amide B and A bands, respectively. Given the presence of the amide bands, the primary content of the membrane-like structure is assumed to be proteins and/or peptidoglycan. Furthermore, the spectrum shows a significant drop in the wavenumber region above 3,280 cm$^{-1}$, which is characteristic of the amide A band. This suggests that the membrane-like structure contains minimal water. If water was present, it would not result in such a significant drop due to the O-H stretching mode, which exhibits a spectral peak at 3,400 cm$^{-1}$, as observed in the background spectrum. In contrast, the intracellular structure appears to contain more water, as its spectrum shows a shoulder at 3,280 cm$^{-1}$, but not a significant drop at higher wavenumbers. This implies that the internal content of the bacterium is an aqueous protein solution. Importantly, these insights cannot be obtained from the morphology alone provided by the OPD image (Fig. 4a), highlighting the significance of MIP contrasts for label-free biochemical analysis.

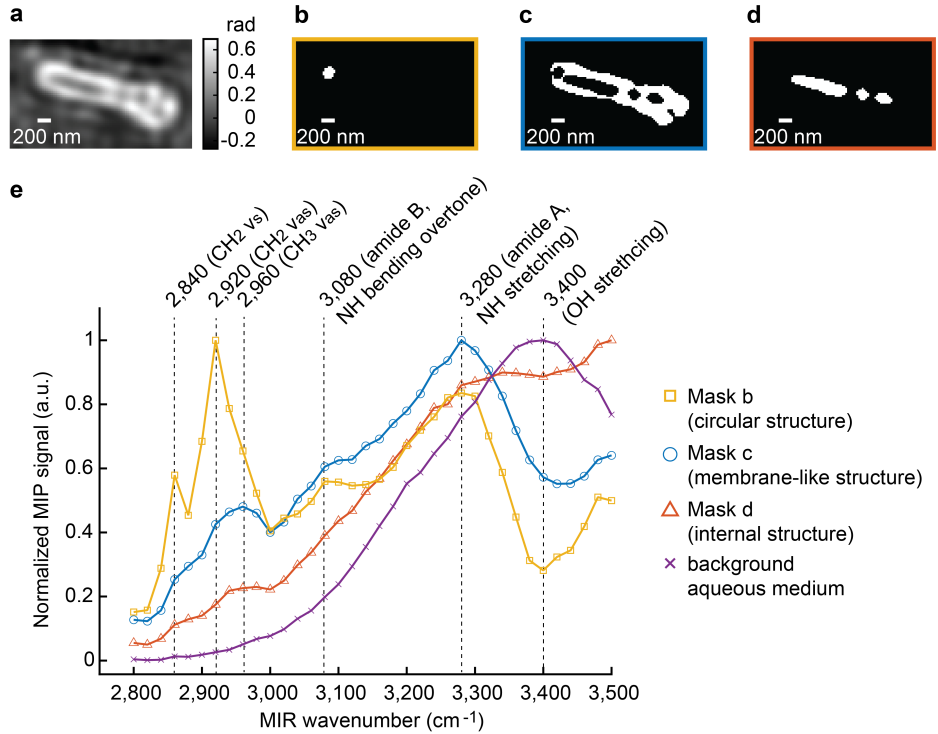

**Figure 4. MIR spectroscopic imaging of bacterial subcellular structures. a** OPD image of a bacterium extracted from commercial fermented soybeans. **b-c** Binary masks for distinguishing subcellular structures. **e** Averaged and normalized MIP spectra obtained within the binary masks shown in **b-d**.

## Discussion

To illustrate the high spatial resolution of our SOSA-MIP-QPI, it is worth comparing the spatial resolution of our method to that of coherent Raman scattering (CRS) microscopy. In CRS imaging, the spatial resolution is enhanced due to their nonlinearity by a factor of $\sqrt{2}$ and $\sqrt{3}$ for stimulated Raman scattering (SRS) and coherent anti-Stokes Raman scattering (CARS) microscopy, respectively. However, these enhancement factors are lower than the maximum achievable gain of 2 through aperture synthesis. Although recent studies have demonstrated super-resolution CRS imaging, e.g., by detecting higher-order (5th- or 7th-order) nonlinear signals, the resolution remains limited to ~200 nm[24]. Generating higher-order nonlinear signals requires a significantly higher flux, which presents challenges for further improvement due to potential sample damage[25]. In contrast, MIP-QPI offers the advantage of improving resolution through aperture synthesis with relative ease. Furthermore, the use of significantly longer pulse durations (~ns) compared to CRS microscopy (~fs-ps) provides the opportunity to employ shorter visible wavelengths to enhance the resolution without causing sample damage, as the occurrence of electronic transition induced by multiphoton processes is relatively infrequent.

The spatial resolution of our system still has room for improvement. Regarding the resolution of synthetic-aperture QPI, which is determined by $\sim \lambda /2(\mathrm{NA}_{\mathrm{illumination}}+\mathrm{NA}_{\mathrm{collection}})$, employing a probe wavelength of 400 nm allows for enhancing the half-pitch spatial resolution to 90 nm with the same synthetic NA of 2.2. Furthermore, by using an oil-based immersion objective lens, the illuminating and collecting NAs can be increased to ~1.4, thereby expanding the synthetic NA up to ~2.8. Collectively, a half-pitch spatial resolution down to 70 nm is within reach. Additionally, regarding the effect of heat diffusion at this high spatial resolution, we can calculate that the resolution degrades from 70 nm to 78, 84, 165, and 480 nm within an aqueous environment for the pulse durations of 0.5, 1, 10, and 100 ns, respectively, based on Eq. (1) in the Methods section. Therefore, 80-nm spatial resolution is still within the range achievable with the pulse duration of 0.5 ns offered by our laser source. The spatial resolution of 100 nm presents a challenging milestone for optical imaging, representing the threshold that distinguishes nanoscopy from microscopy[6]. Therefore, our technique could potentially open the door to far-field MIR nanoscopy.

Finally, compared to the conventional state-of-the-art far-field super-resolution MIR imaging, our SOSA-MIP-QPI achieves an order of magnitude higher imaging speed without compromising spatial resolution. Unlike the higher-order lock-in detection scheme employed in the recently developed techqniue[14], where the probe photons are temporally distributed due to the use of CW light, our SOSA-MIP-QPI utilizes a nanosecond pump-probe measurement scheme. This scheme enables the optimal temporal confinement of the probe photons, allowing them to sense only the non-diffused photothermal signals. As a result, our approach achieves a MIP imaging speed of less than 1 minute/frame, ~60 times faster than the prior art, with a MIP spatial resolution of less than 125 nm. Moreover, By incorporating a faster angle-scanning mechanism of VIS light[26] as well as an image sensor with a higher saturation capacity accompanied by an increased VIS optical power[21], it is possible to achieve an additional order of magnitude improvement in imaging speed.

**Methods**

**Derivation of the optimum pulse duration:** In MIP imaging, the image contrast originates from the photothermal heat induced by the absorption of MIR light by the sample. However, this heat diffuses over time, leading to a degradation of spatial resolution. Therefore, short optical pulse durations are necessary to achieve a spatial resolution as small as the diffraction limit of the VIS imaging system. In this context, we can characterize the size of the photothermal heat[22] as

$$R_T = \sqrt{R^2 + 4kt}, \tag{1}$$

where $k$ is the thermal diffusivity [m$^2$/s], $t$ is the time [s], $R$ is the Gaussian radius [m] of a spherical heat source that provides an impulsive response at $t=0$, and $R_T$ is the Gaussian radius of the photothermal contrast. To ensure that the heat diffusion length is shorter than a certain ratio r of the radius $R$, we need to satisfy the condition

$$R_T = \sqrt{R^2 + 4kt} < (1+r)R, \tag{2}$$

which can be simplified as

$$t < (2r + r^2)R^2/4k. \tag{3}$$

Equation (3) indicates that the requirement on $t$ becomes more stringent as the object diameter $2R$ decreases.

Considering the case of observing an object with a diameter as small as the diffraction limit of a VIS imaging system given by $\Delta x$ [m], we insert $2R = \Delta x$ into Eq. (3) and obtain

$$t < (2r + r^2)\Delta x^2/16k \approx r\Delta x^2/8k. \tag{4}$$

For instance, if we use a VIS microscope with a spatial resolution $\Delta x$=100 nm and allow heat diffusion from an object of this size with r=0.1 in an aqueous sample (k~1.4×10$^{-7}$ m$^2$/s), the required time is t<~0.9 ns. This suggests that both the VIS and MIR light must have pulse durations shorter than 0.9 ns.

**Characterization of the light sources:** A 2-GHz Si photodetector is employed to characterize the temporal profiles of the VIS and the 1,064-nm pulses. Meanwhile, a quantum cascade detector (provided by Hamamatsu Photonics) is used to characterize the temporal profile of the MIR pulse. We used a home-made Fourier-transform infrared spectrometer to measure the spectral profile of the MIR light.

**MIP-QPI:** To acquire MIP-QPI images, the repetition rate of the light sources, the frame rate of the image sensor, and the frequency of the mechanical chopper are synchronized. Specifically, the pulse repetition rate (1 kHz) serves as the master clock of the entire system. The image sensor and the mechanical chopper operate at the 1/14th harmonic (71.4 Hz) and 1/28th harmonic (35.7 Hz) frequencies, respectively. This allows the image sensor to alternately record the MIR-on and -off states. Each raw interferogram is processed to retrieve the complex-field image on a computer. The on-off difference in the phase of the complex-field images is calculated to reveal the MIP contrast. The rotation of the wedge prism is automated using a motorized mount.

Aperture synthesis is performed as described in Supplementary Note 3. A custom-made program is developed to automate the entire measurement process, including rotation of the wedge prism, scanning of the MgO:PPLN crystal, and image acquisition and recording. MIP spectroscopic image acquisition takes ~30 minutes to complete for measuring 31 spectral points between 2,600 and 3,200 cm$^{-1}$ with a spectral resolution of 10 cm$^{-1}$ and a frame-averaging number of 200 (or acquisition of 400 raw images) for one azimuthal angle of illumination.

**Custom-made resolution test chart:** The custom-made spatial-resolution test chart was fabricated using electron-beam lithography. A 520 μm-thick silicon substrate (4-520P1FMM, AKD, Japan) was coated with an electron beam (EB) resist (OEBR-CAN 2.0cp) by spin coating (3000 rpm, 60 s) and baked at 110 °C for 1 minute. The structures were patterned using a rapid EB lithography system (F7000SVD02, Advantest) and then baked at 130 °C for 1 minute. The OEBR-CAN 2.0cp was developed with NMD-W for 1 minute and cleaned with deionized water. Finally, the sample was backed at 90 °C for 90 seconds.

### Data availability
The data presented in this work are available upon request to T. I. with a reasonable reason.

### Acknowledgements
We thank Genki Ishigane for the fruitful discussion. This work was financially supported by Japan Society for the Promotion of Science (20H00125, 23H00273), JST PRESTO (JPMJPR17G2), Precise Measurement Technology Promotion Foundation, Research Foundation for Opto-Science and Technology, Nakatani Foundation, and UTEC-Utokyo FSI Research Grant. Fabrication of the custom-made resolution test chart was performed using the apparatus at the Takeda Clean Room of d.lab at The University of Tokyo.

### Author contributions
M.T. designed the system, carried out the experiment, and analyzed the data. V.R.B. constructed the light sources. H.S. wrote the automated data acquisition program. K.K. fabricated the nanoscale spatial-resolution test chart. M.T. and K.T. performed thermal conduction calculations. T.I. supervised the work. M.T., K.T., and T.I. wrote the manuscript.

### Competing interests
M.T., K.T., and T.I. are inventors of patents related to MIP-QPI.

# Supplementary Information for
# Mid-infrared wide-field nanoscopy

Miu Tamamitsu[1], Keiichiro Toda[1], Venkata Ramaiah Badarla[1], Hiroyuki Shimada[1], Kuniaki Konishi[1], and Takuro Ideguchi[1,*]

[1]Institute for Photon Science and Technology, The University of Tokyo, Tokyo 113-0033, Japan

*ideguchi@ipst.s.u-tokyo.ac.jp

**Supplementary Note 1: Optical system of single-objective synthetic-aperture (SOSA) MIP-QPI**

Figure S1 presents a complete schematic diagram of the optical system. The average input power into the MgO:PPLN crystals is 40 mW. The average power of the VIS pulses after the single-mode polarization-maintaining (PM) fiber is 0.65 mW.

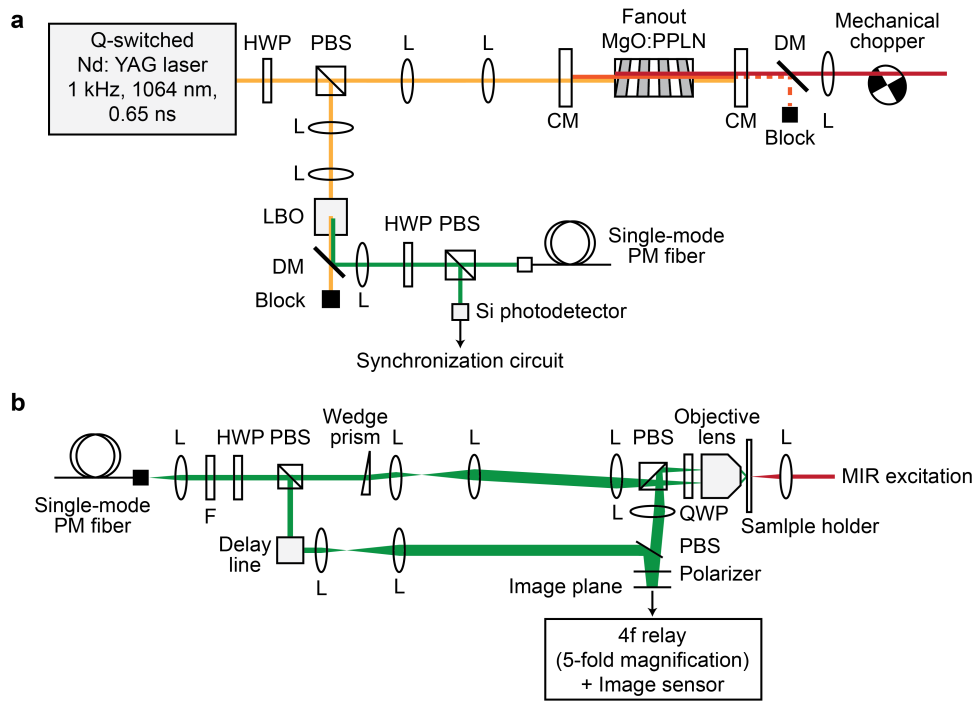

**Figure S1. Complete schematic diagram of the optical system. a** Passively-synchronized sub-ns laser system. **b** Single-objective synthetic-aperture (SOSA) MIP-QPI. HWP: half-wave plate, QWP: quarter-wave plate, PBS: polarizing beamsplitter, L: lens, F: bandpass filter (532 nm, FWHM 4 nm), CM: cavity mirror.

**Supplementary Note 2: Pulse characterization**

Figure S2a shows measured pulse durations of MIR, VIS, and seed laser pulses. Figure S2b shows spectra and pulse energies of MIR pulses at different wavenumbers.

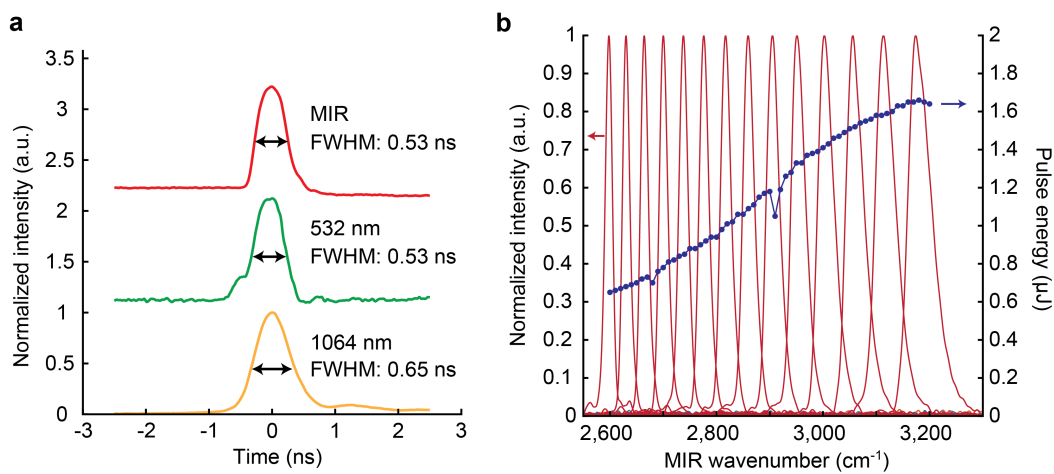

**Figure S2. Pulse characterization. a** Measured pulse durations of MIR, VIS, and seed laser pulses. **b** Spectra and pulse energies of MIR pulses at different wavenumbers.

**Supplementary Note 3: Reconstruction procedure of SOSA-MIP-QPI**

We describe the image-reconstruction procedure employed in our SOSA-MIP-QPI. In addition to the standard synthetic-aperture reconstruction algorithm, SOSA-MIP-QPI requires the computational removal of artifacts caused by MIR-photothermal alteration of the pupil function of the imaging system. Specifically, the MIR light is absorbed not only by the sample but also by the immersion medium between the objective lens and the sample, the cover glass, and/or the glass materials that constitute the objective lens. Such absorption changes the refractive index of these materials, altering the pupil function of the imaging system and thus causing artifacts in the photothermal image. Thanks to the computational nature of synthetic-aperture QPI, this effect can be computationally removed.

In Fig. S3, we use experimental data to explain this effect and the associated image-processing workflow. In this experiment, the custom-made spatial-resolution test chart is excited by MIR light with a wavenumber of 2,600 or 2,920 cm$^{-1}$. The latter wavenumber is resonant with C-H stretching modes in the organic resist, whereas the former is non-resonant.

Using Figs. S3a and S3b, we first explain how the pupil-function alteration induced by the MIR-photothermal effect appears in our raw data. Figure S3a shows an example where the pupil-function alteration is more dominant than the sample-specific MIR photothermal effect. It displays the phase of the differential (i.e., MIR on-off) complex-field image (left panel) and its frequency spectrum (right panel) when 2,600-cm$^{-1}$ MIR light, which is not resonant with the resolution test chart, is irradiated. A localized signal at the center of the circular pupil aperture in the spatial frequency spectrum is evident. This is uncharacteristic of the frequency spectrum of the resolution test chart, where noticeable signals are distributed along the horizontal and vertical directions due to the existence of vertical and horizontal bars, respectively, with various spacings in real space. Indeed, these observations are made with the frequency spectrum and the OPD image of our resolution test chart obtained with standard synthetic-aperture QPI, as shown in Fig. S3b. The non-sample-specific alteration of the pupil function in MIP imaging found in Fig. S3a appears as a non-negligible artifact in real space (left panel in Fig. S3a). Note that since the focused MIR light is irradiated to the sample at the normal incidence, this localized artifact of the pupil appears at its center.

Using Figs. S3c and S3d, we then explain how the effect of the pupil-function alteration is removed during the SOSA-MIP-QPI image-reconstruction procedure. We apply a high-pass filter to the circular pupil of the MIR on-off differential complex-field image so that the non-sample-specific alteration of the pupil is eliminated. The result of this high-pass filtering to the data shown in Fig. S3a is presented in Fig. S3c. In the left panel, we can observe that the non-negligible artifact appearing in real space has vanished due to the high-pass filter. Meanwhile, in the right panel, it is evident that some of the frequency contents are lost due to the high-pass filter. However, thanks to the aperture synthesis, the lost information can be filled with the measurements performed at other illumination angles. In an example shown in Figure S3d, the use of 8 azimuthal angles of illumination allows for filling the lost regions in the final synthetic aperture of the MIR on-off differential image. In the created synthetic spectrum, there is no evident signal with the non-resonant 2,600-cm$^{-1}$ light turned on, whereas strong signal distribution in the horizontal and vertical directions, which is characteristic of the resolution test chart (Fig. S3a), can be observed with the resonant 2,920-cm$^{-1}$ turned on.

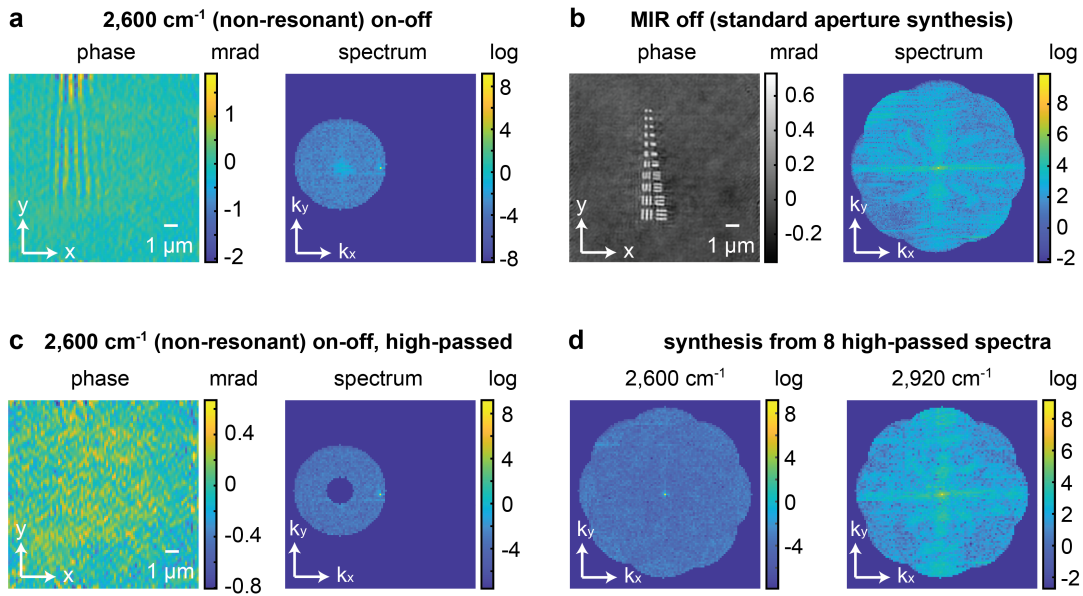

**Figure S3. Reconstruction procedure of SOSA- MIP-QPI. a** MIP image of the spatial-resolution test chart (left panel) and its frequency spectrum (right panel), obtained when the MIR light at 2,600 cm$^{-1}$ is irradiated. This wavenumber is not resonant with the vibrational modes included in the material of the test chart. A localized signal appears at the center of the circular pupil aperture in the spatial-frequency spectrum (right panel), which does not reflect the sample-specific signal distribution observed in the right panel of **b** and hence results in a non-negligible artifact in the real-space OPD image (left panel). **b** OPD image of the spatial resolution test chart (left panel) and its spatial frequency spectrum (right panel) obtained by standard aperture synthesis with the MIR light turned off. **c** Similar to **a**, but after the artifact appearing at the center of the circular pupil aperture has been removed through computational high-pass filtering (right panel). It results in the removal of the artifact in the real-space image (left

panel). **d** Synthetic-aperture spectra obtained with 8 azimuthal angles of illumination. The central empty region of the computational aperture shown in the right panel of **c** is filled using these 8 azimuthal angles. When the non-resonant 2,600-cm$^{-1}$ MIR light is irradiated, there is no evident signal in the synthetic spectrum. In contrast, when the 2,920-cm$^{-1}$ MIR light, which is resonant with the C-H stretching mode included in the organic resist, is irradiated, strong sample-specific signals appear in the synthetic spectrum, showing similar horizontal and vertical distributions to those observed in the OPD image of the resolution test chart (right panel of **b**).

**Supplementary Note 4: Custom-made nanoscale spatial-resolution test chart**

Figure S4 shows a scanning electron micrograph of the custom-made nanoscale spatial-resolution test chart. Several such charts were fabricated on the same silicon substrate. In the image, the 122-nm half-pitch bars are not spaced properly, but for the experiments, a chart is selected that has well-spaced 122-nm half-pitch bars.

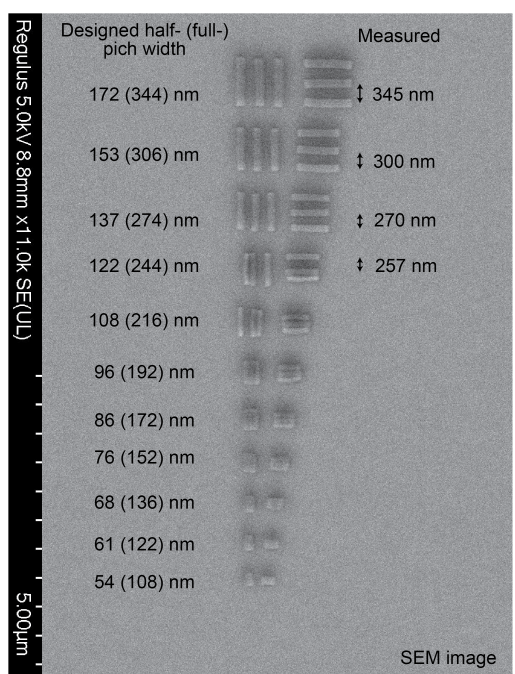

**Figure S4. Scanning electron micrograph of the custom-made nanoscale spatial-resolution test chart.**

**Supplementary Note 5: Principal component analysis of MIP spectroscopic images of the bacterium**

Figure S5 shows the result of the principal component analysis (PCA) performed on the MIP spectral image dataset presented in Fig. 4 in the main text. The first three principal components (PCs) are found to represent noticeable bacterial subcellular structures. From this result, we estimated the three binary masks (Fig. 4b-d in the main text) that represented these subcellular structures and used them for a more detailed MIR spectroscopic analysis presented in the main text.

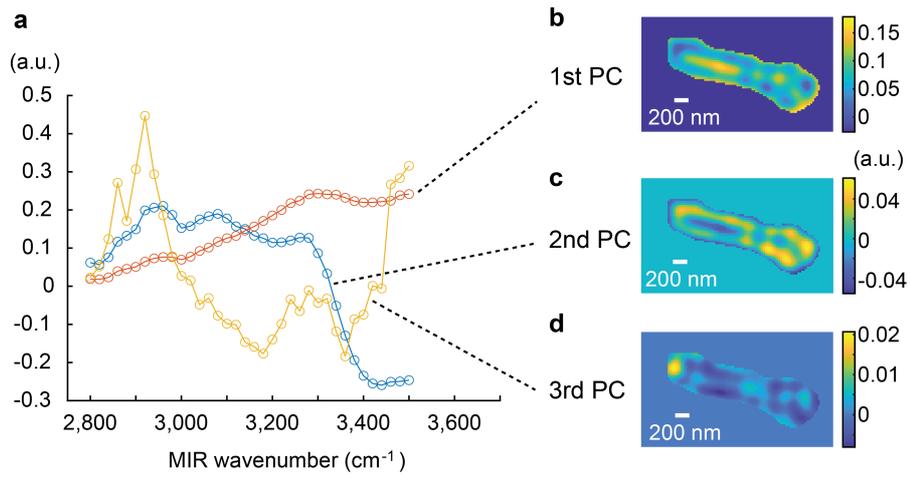

**Figure S5. PCA of the MIP spectroscopic images of the bacterium. a** Coefficients of the first three PCs, representing unique MIR spectral signatures. **b-d** Score images of the first three PCs, representing unique subcellular structures.